# Service Aware Fuzzy Logic Based Handover Decision in Heterogeneous Wireless Networks

**Abstract:** The ubiquitous services of wireless communication networks are growing rapidly by the development of wireless communication technology. While a user is roaming from one cell to another cell an intelligent decision mechanism and network selection is extremely needed to maintain the quality of service (QoS) during handover. Handover decision must be made precisely to avoid many unnecessary obstacles like ping-pong, corner effect, shadow effect, call blocking, and call dropping probability etc. This work focused on services like voice, video, and data during handover decision using fuzzy logic in heterogeneous network environment. Service is an important factor for the users and particular services require respective QoS. In this paper we provided all the cases of handover decisions between macrocell and femtocell networks considering service type. The proposed system models regarding this handover decision using fuzzy logic considering several input parameters e.g. received signal strength indicator (RSSI), data rate, user's velocity, and interference level (signal-to-noise plus interference ratio) to make handover from femtocell to macrocell, macrocell to femtocell or femtocell to femtocell. The performance of different parameters are shown based on service type are analyzed.

## I. INTRODUCTION

Next generation wireless communication needs seamless connectivity with better quality of service (QoS), high data rate services, favorable price, and multimedia applications among different access networks. Heterogeneous networks are the combination of different networks which can provide different services and quality with verities of features. So handover is a common criterion in heterogeneous network and efficient handover decision is an important issue [1]. The mechanism of transferring an ongoing call from one cell to another or a mobile user switching from one network to another is called handover process. Handover is one of the challenging issues at present as communication is progressing from 4G to 5G [2]. There are many wireless technologies which require interconnection like macrocell, femtocell, WLAN, and WiMAX. Handover decision is to determine the best expected access network and decide at any particular time whether to carry out handover or not [3], [4]. Network selection is one of the major issues, because without selecting proper network our purposes of handover are not fulfilled. Macrocell can support high mobility whereas femtocell cannot support high mobility. We know different services need different QoS parameters for preference. Some services prefer macrocell with better QoS level on the other hand, some favor femtocell, since femtocell cannot support high mobility but femtocell can support high data rate and throughput. Femtocellular technology is widely deployed in subscribers' homes to provide high data rate communications with better QoS [5]. The femto-access-points (FAPs) enhance the service quality for the indoor mobile users. Some key advantages of femtocellular network technology are the improved coverage, reduced infrastructure and the capital costs, low power consumption, improved signal-to-noise plus interference ratio (SNIR) level at the mobile station (MS), and improved throughput. Femtocells operate in the spectrum licensed for cellular service providers [6], [7].

In the past, a few literatures about fuzzy-based solution for vertical handover decision systems have been proposed [8], [9]. A fuzzy-based vertical handover decision algorithm which assumes interconnection between WLAN and WMAN is proposed in [10]. The decision parameters considered here are: received signal strength indicator (RSSI), data rate, usage cost, and user preference. Previous researchers considered only one case (one direction of handover) and handover occurred from WMAN to WLAN but they did not consider user's velocity and services during handover decision. In a more recent work, handover decision using a Kalman filter and fuzzy logic in heterogeneous wireless networks has been shown handover decision from cellular networks to WLAN [11]. In our research, we consider three different handover scenarios, e.g. femtocell to femtocell, femtocell to macrocell, and macrocell to femtocell. Femtocell is small coverage area so it's tolerance of high mobility is less than macrocell. Therefore, we included velocity as a handover selection parameter for both networks. This research also emphasizes on service type as handover decision parameters. In the past, researcher did not consider service type for handover decision. Previous researches only consider one direction of handover but we consider multi directions of handover.

The rest of this paper is organized as follows. Section II shows handover decision in details including scenario and call flow of system model and fuzzy controller of handover decision. Performance analysis and evaluation of services for different network parameter considering different conditions are shown in Section III. Finally Section IV contains the concluding notes of work summary.

## II. HANDOVER DECISION

### A. Path Loss Model for Channel and Network selection.

Here we adopt path loss channel propagation model for macrocell users can be expressed as [7]:

$$L = 69.55 + 26.16 \log_{10} f_{c,m} - 13.82 \log_{10}(h_b) - a(h_m) + [44.9 - 6.55 \log_{10} h_b] \log_{10} d + L_{sh} \quad [dB] \quad (1)$$

$$a(h_m) = 1.1[\log_{10} f_{c,m} - 0.7]h_m - (1.56 \log_{10} f_{c,m} - 0.8) \quad (2)$$

where $L$ is the path loss, $f_{c,m}$ is the center frequency in MHz of the macrocell, $h_b$ is the height of the macrocellular BS in meter, $h_m$ is the height of the MS in meter, $d$ is the distance between the macrocellular BS and the MS in kilometer, $L_{sh}$ is the shadowing standard deviation.

The propagation model for femtocell users can be expressed as [7]:

$$L_{femto} = 20\log_{10} f_{c,f} + N\log_{10} d_1 - 28 [dB] \quad (3)$$

where $f_{c,f}$ is the center frequency in MHz of the femtocell, $d_1$ is the distance between the FAP and the MS in meter.

The expression of RSSI is

$$P_R = P_T 10^{\frac{-L}{10}} \quad (4)$$

where $P_T$ is the transmitted power and $P_R$ is the received power, and L is the path loss.

The received SNIR level of femtocell user in a macrocellular integrated network or femtocellular integrated network can be expressed as

$$SNIR = \frac{S_{f0}}{\sum_{i=0}^{N} I_{i,f} + \sum_{p=0}^{M} I_{i,m} + N_l} \quad (5)$$

where $S_{f0}$ is the power of the received signals from the associated macrocellular base station or FAP, $I_{i,f}$ is the power of the interference signal from the i-th interfering femtocell from among the $N$ neighboring femtocell users and $I_{i,m}$ is the received interference signal from the p-th macrocell from among the $M$ macrocell users. $N_l$ is the presented noise level.

The data rate of femtocell and macrocell users achieved are,

$$C_{f(m)} = B_{f(m)} \log_2 \left(1 + SNIR_{f(m)}\right) \quad (6)$$

where $B_{f(m)}$ is the allocated bandwidth for a user in femtocell (macrocell) network and $SNIR_{f(m)}$ is the SNIR level at targeted femtocell(macrocell) networks.

### B. Scenarios of System Model

In this paper, we consider two scenarios of the system models which are shown in Fig. 1. First scenario is for case-1 in which the MS is situated inside macrocell and handover occurs either to target femtocell network or remains in macrocell. Second scenario is for case-2 in which the MS is situated inside the femtocell. Two options are possible for these cases where handover occurs either from femtocell to femtocell or femtocell to macrocell. We also consider three service types such as voice, video, and data for calculating handover factor. The availability of the network space is indicated by RSSI level. Data rate is considered by available throughput of both networks. Velocity is indicated by user's mobility in which MS travels within the network boundary. Interference considers the quality of signal level and noise level where user achieved from both macrocell and femtocell networks.

### C. Call flow of System Model

Fig. 2 shows the basic call flow for case-1 scenario of the system model. The MS measures RSSI of macro base station.

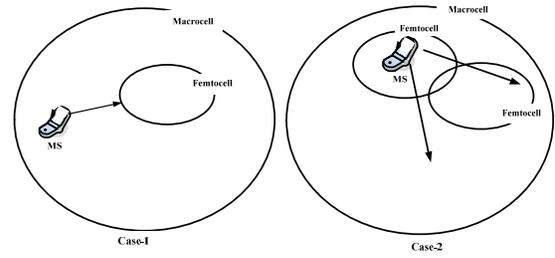

**Fig. 1** Different handover scenarios for macrocell/femtocell integrated network

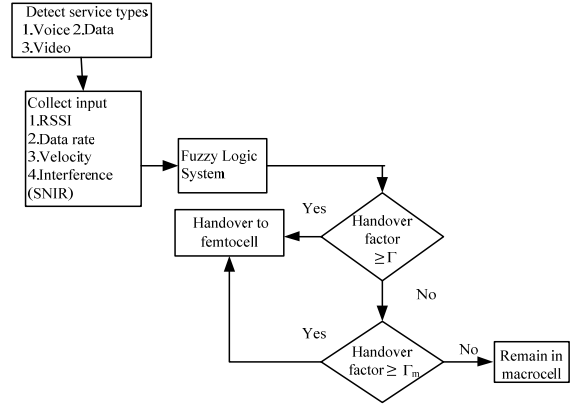

**Fig. 2** System model for case-1 handover decision.

In our proposed model we consider two networks one is femtocell and another is macrocell. First the system detects the service type the mobile user required to execute then after detecting the service type the system collects the input as RSSI, data rate, velocity, and SNIR. All inputs are combined then feed to the fuzzy logic system for processing the mechanism of handover factor. Here $\Gamma$ is the threshold factor value to make handover decision to femtocell network. $\Gamma_m$ is the calculated handover factor of macrocell. If the handover factor is greater or equal to threshold factor $\Gamma$ then MS initiates handover to target femtocell networks otherwise if not then MS goes to next condition. If handover factor is greater or equal to $\Gamma_m$ then handover to femtocell otherwise MS remains in the current macrocell network.

Fig. 3 shows the basic call flow for case-2 scenario of the system model. $\Gamma_f$ is the calculated handover factor for target femtocell. Firstly the system feeds the input parameters based on service type to fuzzy sub-system. Then it calculates the handover factors for two different target networks, one is for femtocell network and another for macrocell network. After calculating two handover factors, it makes a decision by comparing two different handover factors. We add a weight value $K$ with femtocell's handover factor. If the value of $K$ increases then the preference for femtocell networks will increases for handover. Therefore, the operator has the control to change the preference of macrocell or femtocell networks on the basis of requirement. If $\Gamma_m$ is not larger than $\Gamma_f$ then it decides for handover to femtocell network otherwise it goes for the next condition. Here if $K\Gamma_f$ is larger or equal to $\Gamma_m$ then it decides for handover to femtocell also otherwise it makes handover to macrocell network. Finally decision to handover will be successfully accomplished in our system model.

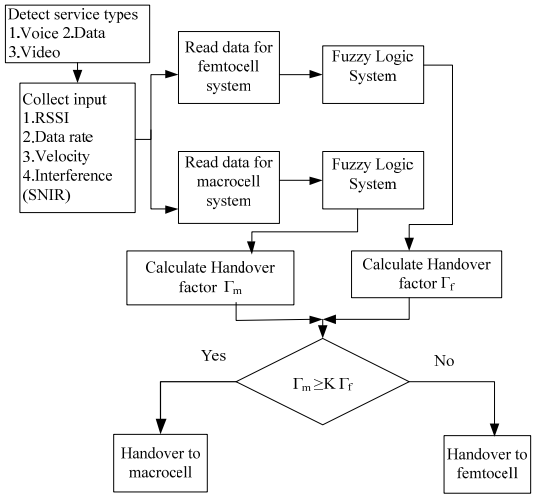

**Fig. 3** System model for case-2 handover decision.

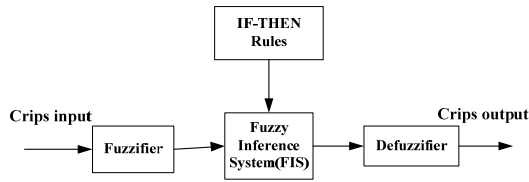

**Fig. 4** Basic architecture of Fuzzy logic system.

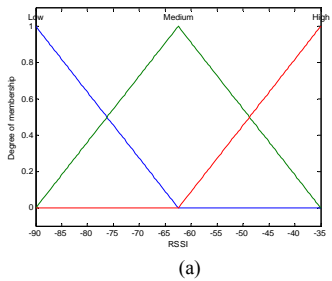
(a)

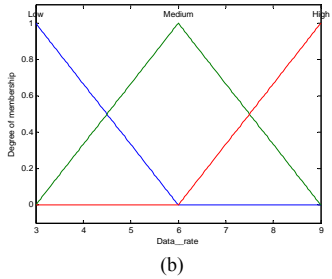
(b)

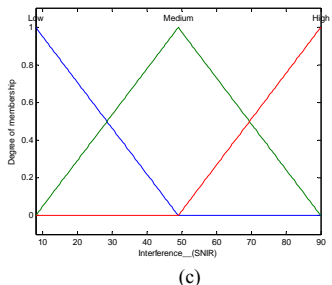
(c)

**Fig. 5** Membership function of the inputs considering user (a) RSSI, (b) data rate, and (c) interference level (SNIR) for voice service

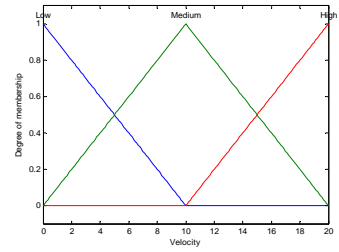
(a)

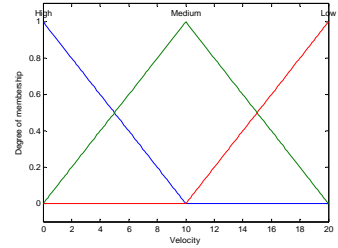
(b)

**Fig. 6** Membership function of the inputs considering user velocity in (a) macrocell and (b) femtocell networks for voice service

### D. Controller of Handover Decision

The basic architecture of a fuzzy logic system is shown in Fig. 4. It consists of four components fuzzifier, IF-THEN rules, fuzzy inference system (FIS), and defuzzifier. We use a Mamdani Fuzzy inference system (FIS) that is composed of the functional blocks. Each of the fuzzy sets has four inputs (RSSI, data rate, velocity, and SNIR) and three membership functions (MF) of Low, Medium, and High. Here we choose triangular MF for simple, smooth, and absolute value at all.

Fig. 5 shows the MF of the networks input parameters as (a) RSSI for both macrocell and femtocell users ,(b) Data rate for both macrocell and femtocell users, and (c) Inerference level(SNIR) for both macrocell and femtocell users for voice service. Fig. 6 shows the MF of neworks parameters as (a) velocity for macrocell user (b) velocity for femocell user for voice service.The input of network parameters are divided into three catagories low, medium, and high levels. The membership funcion of voice service is shown only graphically. The input parameters of the MF for video are shown in table along with voice service. Table 1 shows the inputs for voice user and video user for both macrocell and femtocell networks. The ranges of inputs for fuzzy variables RSSI, data rate, velocity, SNIR are also shown in Table 1.

Fig. 7 shows the MF of handover factor, it is divided into six catagories lower, low, lower medium, higher medium, high, and higher level. The Fuzzy system has rules base contains IF-THEN rules, which required by the fuzzy inference system. There are several antecedents that are combined using fuzzy operators such as AND, OR, and NOT. Here we designed four fuzzy inputs variables and three fuzzy sets for each fuzzy variable, hence the maximum possible number of rules in our rule base is $3^4=81$.The fuzzy output decision sets are arranged into a single fuzzy set and passed through the defuzzifier to be converted into precise quantity, the handover factor, which determines whether a handover is necessary or not as still

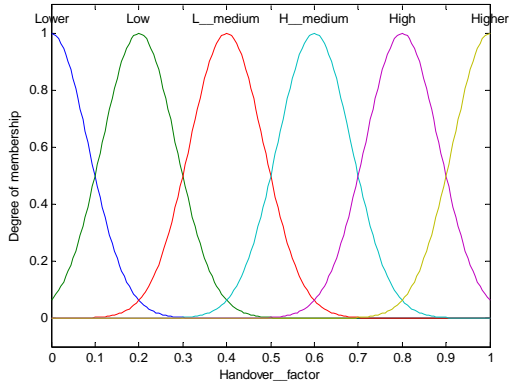

**Fig. 7** Memberships function of handover factor

**Table 1** Ranges of inputs (voice & video user)

|  | Inputs | Low | High | Units |
|---|---|---|---|---|
| Voice user | RSSI (macro and femto) | -90 | -35 | dBm |
|  | Data rate (macro and femto) | 1 | 4 and 7 | Mbps |
|  | Velocity (macro) | 0 | 20 | km/hr |
|  | Velocity (femto) | 20 | 0 | km/hr |
|  | Interference level (SNIR) (macro and femto) | 8 | 90 | dB |
| Video user | RSSI (macro) | -62.5 | -7.5 | dBm |
|  | Data rate (macro) | 3 | 9 | Mbps |
|  | Velocity (macro) | 0 | 20 | km/hr |
|  | Interference level (SNIR) (macro) | 49 | 131 | dB |
|  | RSSI (femto) | -117.5 | -62.5 | dBm |
|  | Data rate (femto) | 1 | 6 | Mbps |
|  | Velocity (femto) | 20 | 0 | km/hr |
|  | Interference level (SNIR) (femto) | -33 | 49 | dB |

remain the same network. The range of handover factor is from 0 to 1 which is a Gaussian function shown in Fig.7. The maximum membership of the sets Lower and Higher at 0 or 1, respectively. We show some of the rules among 81 IF-THEN rules below,

- Rule-1: If RSSI is low and Data rate is low and Velocity is low and SNIR is low then handover factor is lower.
- Rule-25: If RSSI is low and Data rate is high and Velocity is high and SNIR is low then handover factor is lower medium.
- Rule-50: If RSSI is medium and Data rate is high and Velocity is medium and SNIR is medium then handover factor is high.
- Rule-81: If RSSI is high and Data rate is high and Velocity is high and SNIR is high then handover factor is higher.

### III. PERFORMANCE ANALYSIS

In this section, we analyze the performance of our proposed scheme. We considered voice and video services for the analysis. We can easily make decision to switch between macrocell and femtocell. Figs. 8-10 show the calculated handover factors with respect to velocity of users considering both macrocell and femtocell as target networks. We considered different conditions of network parameters as RSSI, data rate, and SNIR. We assume *K=1* for the analysis.

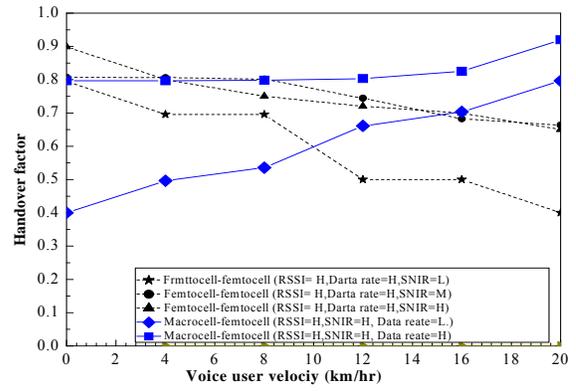

(a)

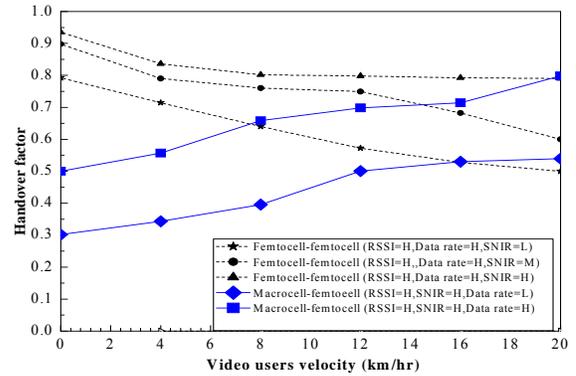

(b)

**Fig. 8** Handover factor vs user velocity considering at high inputs parameters (a) voice service and (b) video service

The system calculates the handover factors and then based on the calculated factors; the target network for handover is selected.

Figs. 8-10 also show femtocell and macrocell networks handover factors for voice and video services. In Fig. 8, RSSI of macrocell user are at high condition. In femtocell network, user's RSSI and data rate are high condition but varied SNIR as low, medium, and high conditions. On the other hand, in macrocell network, user's SNIR are kept high but varied data rate as low, medium, and high conditions. In case of voice both (femto and macro) curves intersect at velocity 10 km/hr and 16 km/hr. Before these velocity points, the MS can choose femtocell. However, after these intersecting points, for case-1 scenario, the MS handover to femtocell or remains in macrocell network as RSSI and SNIR are high. For case-2 scenario, the MS choose for handover either macrocell or femtocell. Another case is video, at the same condition of input parameters at voice the MS choose femtocell network most. There are two intersecting points at velocity 8.5 km/h and 16.5 km/hr and after the intersecting points, for case-1 scenario, the MS remains in macrocell or choose femtocell network if necessary. However, for case-2 scenario, the MS choose handover to femtocell or macrocell networks which is more preferable for this service but handover factor is higher in femtocell for video so it is chosen at this situation. At the same conditions of network parameters, the value of handover factor for voice service differs from handover factor of video service. Handover

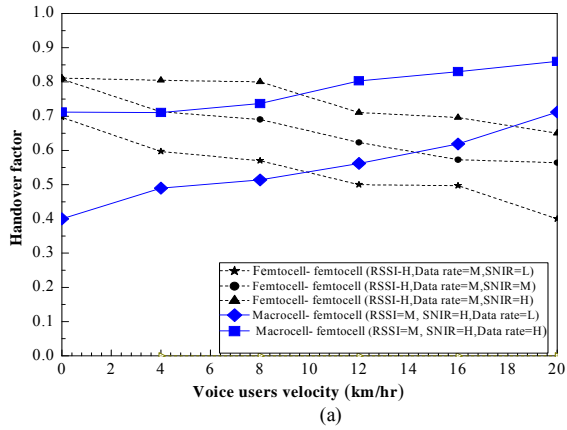
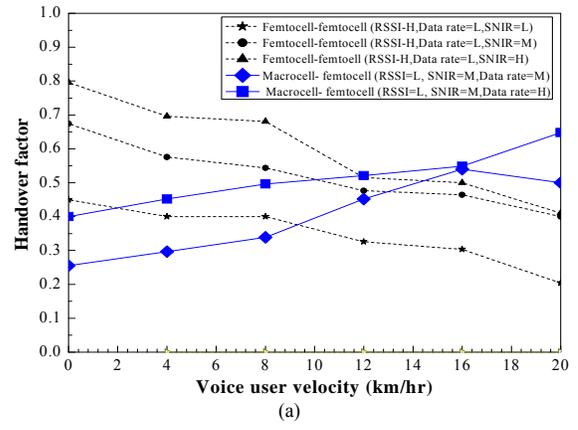

(a)

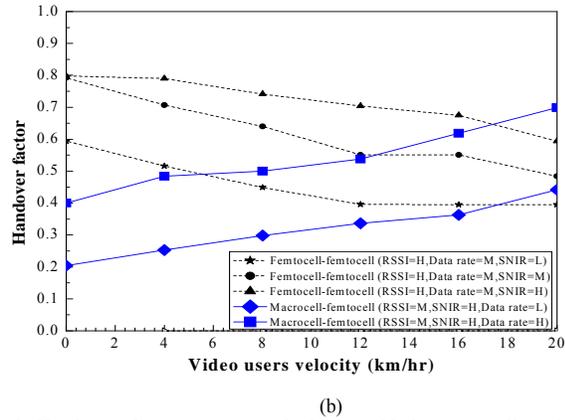
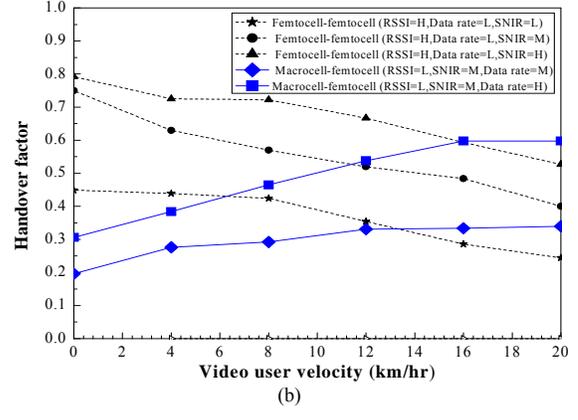

(b)

**Fig. 9** Handover factor vs user velocity considering at medium inputs parameters (a) voice service and (b) video service

**Fig. 10** Handover factor vs user velocity considering at low inputs parameters (a) voice service and (b) video service

factor of voice at femtocell is less but more in video. For user consideration, choice and availability of networks parameters some facts femtocell are preferable and for some facts are macrocell. For video, femtocell is more preferable than macrocell which is shown in Fig. 8. In Fig. 9 RSSI of microcell user's are at medium conditions. In femtocell network, user's RSSI are high conditions and data rate are medium condition. However, user's SNIR varied as low, medium, and high conditions. On the other hand, in macrocell network, user's SNIR are high but varied data rate as low, medium and high conditions as well. In case of voice both (femto and macro) curves intersect velocity at 4 km/hr and 10 km/hr, and 14 km/hr. Before these velocity points, the MS can choose femtocell. However, after passing intersecting points, for case-1 scenario, MS choose femotocell for handover or remain in macrocell as RSSI and SNIR are medium. for case-2 scenario, MS choose handover either macrocell or femtocell as the availability of networks parameters. Another case is video, but at the same conditions for video, velocity shift from 5km/hr, 13km/hr, and 18km/hr, respectively and MS choose femtocell network most that's why it's velocity is shifted and more time to stay in femtocell network. There are three intersecting points at velocity 5 km/h and 13 km/hr and 18 km/hr after the intersecting points the MS choose macrocell if necessary but handover or stay to femtocell is much more preferable as handover factor is higher in femtocell for video. At the same conditions of network parameters, the value of handover factor for voice service differs from handover factor of video service. Handover factor of voice at femtocell is less but more in video.

For user consideration, choice and availability of networks parameters some facts femtocell are preferable and for some facts are macrocell. Here in case of video, femtocell is more acceptable. However, more intersecting points give MS an opportunity to choose between macrocell and femtocell for handover considering user preference. In Fig. 10 RSSI of macrocell user's are at low condition. In femtocell network, user's RSSI are high condition and data rate are low condition but varied SNIR as low, medium, and high conditions. On the other hand, in macrocell network, user's SNIR are medium condition but varied data rate as low, medium and high conditions. In case of voice both (femto and macro) curves intersect velocity at 2 km/hr and 9 km/hr, and 12 km/hr and, 16 km/hr. Before these velocity points, the MS can choose femtocell or remain in macrocell network, for case-1 scenario. However, after passing intersecting points above velocity condition of handover, for case-2 scenario, MS can choose handover either macrocell or femtocell. Another case is video, but at the same conditions for video, velocity shift from 7 km/hr, 12.5 km/hr, and 16.5 km/hr, respectively and MS choose femtocell network most that's why it's velocity is shifted and more time to stay in femtocell network. There are three intersecting points at velocity 7 km/h and 12.5 km/hr and 16km/hr after the intersecting points the MS choose macrocell if necessary but handover or stay to femtocell is much more preferable as handover factor is higher in femtocell for video. At the same conditions of network parameters, the value of handover factor for voice service differs from handover factor of video service. Handover factor of voice at femtocell is less but more in video. For user consideration, choice and availability of networks parameters some facts femtocell are preferable and for some facts are macrocell. Here in case of video, femtocell is more as well. From the overall analysis, we can conclude that video service prefers to handover femtocell more than macrocell

The designs are simulated using fuzzy logic tool on MATLAB platform. In our simulation, we consider two services which are badly needed in mobile communication. In case of video and data services we can prefer femtocell more than macrocell. So by considering the perspective of services of users the proposed model will surely help the distribution of valuable spectrum.

## IV. CONCLUSION

There is an increase in the percentage usage of many high data rate applications over the past few years and this trend will most certainly persist on in the future. Due to the high demand, our wireless technologies such as WiMAX, WiFi, WLAN, femtocell, macroccell etc. have been researched upon to improve user types and prerequisite. Our work mainly designs a handover decision mechanism. The handover design mechanism is considered for macrocell/femtocell integrated network. We consider macrocell to femtocell, femtocell to macrocell and femtocell to femtocell all three possible handover scenarios. Different service requires different QoS level. If the system uses same handover process for all cases of services then smooth and efficient handover may not be executed. For this concern, we consider service priority handover call. Therefore, we calculate handover factor and based on the handover factor the handover decision is executed. The research proposes service aware fuzzy rule-based intelligent handover decision where network parameters are RSSI, data rate, velocity and interference level (SNIR). We analyze the performance of network parameters considering service type such as voice, video, and data. By investigating performance parameter of velocity vs handover factor, we notice that by the decreasing of input parameter the intersecting points are increasing in this case the value of network parameter are low which affect the macrocell and femtocell networks. Our proposed scheme shows that femtocell is preferable for video service as it requires high data rate and low cost. Voice prefers macrocell as it needs high RSSI, high mobility and less delay. If a situation is created that handover factor of femtocell is greater than handover factor of macrocell or the performance curve of macrocell is crossed higher than femtocell then MS chooses femtocell for handover as it is low cost and support high data rate and less traffic load. The analysis also indicates the different effects of macrocell and femtocell networks for different conditions of network parameters. However, our proposed scheme provides a good basis for research of handover decision using fuzzy logic based on user services successfully.

## REFERENCES

[1] A. Calhan and C. Ceken, "Case Study on Handoff Strategies for Wireless Overlay Networks," *Computer Standard Interfaces*, vol. 35, no. 1, pp. 170–178, January 2013.

[2] R. Khan, S. Aissa, and C. Despins, "Seamless Vertical Handoff Algorithm for Heterogeneous Wireless Networks—An Advanced Filtering Approach," in Proceeding of *IEEE Symposium on Computers and Communications*, July 2009, p. 255-260.

[3] M. Zekri and B. Jouaber, "A Review on Mobility Management and Vertical Handover Solutions Over Heterogeneous Wireless Networks," *Journal of Computer Communications*, vol. 35, no. 17, pp. 2055-2068, October 2012.

[4] S. Wang and C. Fan, "A Vertical Handoff Method via Self-Selection Decision Tree for Internet of Vehicles," *IEEE Systems Journal*, pp. 1-10, March 2014.

[5] M. Z. Chowdhury and Y. M. Jang, "Handover Management in High-dense Femtocellular Networks," *EURASIP Journal on Wireless Communications and Networking*, pp.1-21, January 2013.

[6] K. N. Azam and M. Z. Chowdhury, "Intelligent Interference Management Based on On-Demand Service Connectivity for Femtocellular Networks," in Proceeding of *International Conference on Informatics, Electronics & Vision (ICIEV)*, May 2013, p. 1-6.

[7] M. Z. Chowdhury, N.Saha, S.H. Chae, and Y. M. Jang, "Handover Call Admission Control for Mobile Femtocells with Free-Space Optical and Macrocellular Backbone Networks," *Journal of Advanced Smart Convergence*, vol. 1, no. 1, pp. 19-26, May 2012.

[8] T. Thumthawatworn, A. Pervez, and P. Santiprabhob, "Adaptive Modular Fuzzy-based Handover Decision System for Heterogeneous Wireless Networks," *International Journal of Networks and Communications*, vol. 3, no. 1, pp. 25-38, May 2013.

[9] A. Calhan and C. Ceken, "An Optimum Vertical Handoff Decision Algorithm Based on Adaptive Fuzzy Logic and Genetic Algorithm," *Wireless Personal Communications*, vol. 64, no. 4, pp. 647-664, June 2012.

[10] Q. He, "A Fuzzy Logic Based Vertical Handoff Decision Algorithm between WWAN and WLAN," in Proceeding of *International Conference on Networking and Digital Society,* June 2010, p. 561–564.

[11] L. kustiawan and K. H. Chi, "Handoff Decision Using a Kalman Filter and Fuzzy Logic in the Heterogeneous Wireless Networks," *IEEE Communications Letters,* vol. 19, no. 12, pp. 2258-2261, December 2015.